\documentclass{article}
\usepackage{spconf, amsmath, epsfig, amssymb, booktabs, multirow, soul, color, xcolor}
\usepackage{graphicx}
\usepackage{hyperref}
\hypersetup{
    colorlinks=true,
    linkcolor=blue,
    urlcolor=red,
    citecolor=blue,
}

\title{Arctic Sea Ice Image Super-Resolution Based on Multi-Scale Convolution \\ and Dual-Gating Mechanism}

\name{Zhaomin Fang$^{1}$, Wankun Chen$^{1}$, Feng Gao$^{1}$, Yanhai Gan$^{1}$, Junyu Dong$^{1}$, Yang Zhou$^2$}
\address{$^1$ School of Computer Science and Technology, Ocean University of China, Qingdao 266100, China \\
$^2$ China Electronic Standardization Institute Huadong Branch, Suzhou 215124, China
\thanks{This work was supported in part by the National Key Research and Development Program of China under Grant 2022ZD0117202 and in part by the Natural Science Foundation of Qingdao under Grant 23-2-1-222-ZYYD-JCH. (Corresponding author: Feng Gao, Email: gaofeng@ouc.edu.cn)}}

\begin{document}
\maketitle

\begin{abstract}

Arctic Sea Ice Concentration (SIC) is the ratio of ice-covered area to the total sea area of the Arctic Ocean, which is a key indicator for maritime activities. Nowadays, we often use passive microwave images to display SIC, but it has low spatial resolution, and most of the existing super-resolution methods of Arctic SIC don't take the integration of spatial and channel features into account and can't effectively integrate the multi-scale feature. To overcome the aforementioned issues, we propose MFM-Net for Arctic SIC super-resolution, which concurrently aggregates multi-scale information while integrating spatial and channel features. Extensive experiments on Arctic SIC dataset from the AMSR-E/AMSR-2 SIC DT-ASI products from Ocean University of China validate the effectiveness of porposed MFM-Net.

\end{abstract}


\section{Introduction}

The Arctic sea ice plays an important role in maintaining the Earth's climate system. It can provide guidance for many maritime activities such as fishing and transportation \cite{xian2017super}. Sea ice concentration (SIC) is the ratio of the area covered by sea ice to the total area of sea, which can intuitively indicate changes in sea ice.

SIC can be computed by performing image processing on different types of remote sensing images, such as synthetic aperture radar (SAR) images, optical satellite images,  and passive microwave images \cite{shen2021comparison}. Passive microwave images are generated by detecting microwave radiation emitted naturally by the Earth’s surface, which have a wide coverage range, strong penetration, and high temporal resolution, making them an important data source for the continuous monitoring of the Arctic SIC. The Advanced Microwave Scanning Radiometer for EOS (AMSR-E) is one of the representative datasets of passive microwave images. However, due to the limitations of observation and imaging equipment, the spatial resolution of passive microwave images is only about 12.5–50km,  which is adverse to the more precise sea ice monitoring \cite{shen2021comparison}. Therefore, generating high-resolution images of Arctic SIC from passive microwave images is still an important and meaningful topic for monitoring the Arctic environment.

Recently, many deep learning methods have been proposed to improve the performance of Arctic sea ice image super-resolution, such as the encoder-decoder network \cite{feng2022super}, PMDRnet \cite{liu2022pmdrnet}, CNN-based method \cite{cooke2019estimating} and OGSRN \cite{yanshan2022ogsrn}. All of these methods have achieved good performance on Arctic sea ice data super-resolution, but these method can hardly integrate the multi-scale information and ignore the integration of the spatial and channel features.

To solve the these problems, we propose a novel Multi-scale Feature Modulation Network (MFM-Net) to improve the spatial resolution of passive microwave images of Arctic SIC. To be more specific, we design a Multi-Scale feature Fusion Module (MSFM) to dynamically select representative features. In addition, we design a Dual-Attention Gated Module for non-linear feature transformation. To demonstrate the effectiveness of the proposed MFM-Net, we conduct extensive experiments on the AMSR SIC data. The experimental results show that the proposed MFM-Net performs better than state-of-the-art methods.

\begin{figure*}[h]
\centering
\includegraphics [width=6.5in]{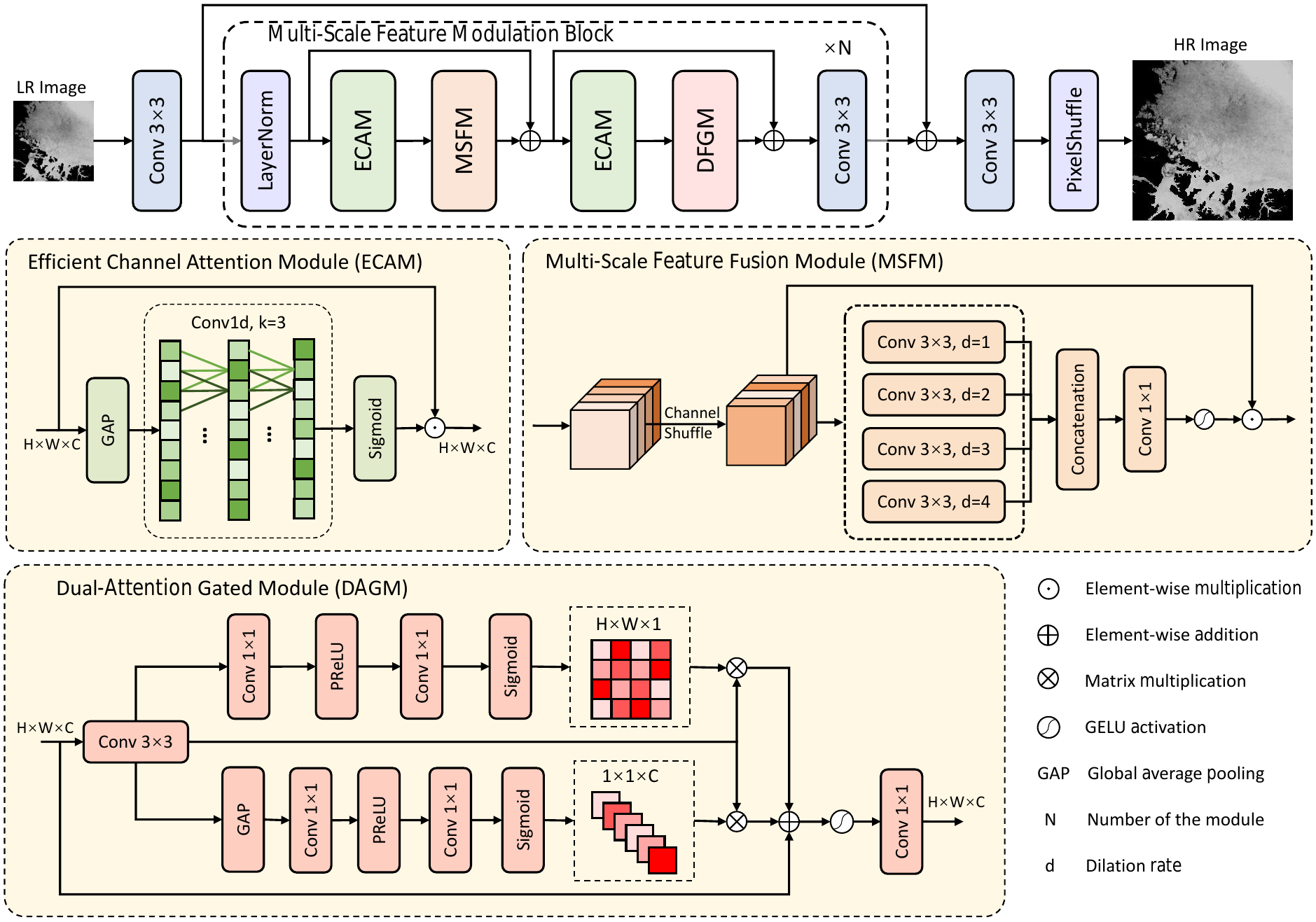}
\caption{Overall architecture of the proposed MFM-Net for Arctic SIC image super-resolution.}
\label{fig_1} 
\end{figure*}

\section{Methodology}

The framework of the proposed MFM-Net is illustrated in Fig. \ref{fig_1}. MFM-Net is comprised of two parts: a stack of multi-scale feature modulation blocks, and an image reconstruction block. 

Given a low-resolution SIC image $I_{LR} \in \mathbb{R}^{H \times W \times 1}$, we firstly use a $3 \times 3$ convolution layer for shallow feature extraction, so that we can generate the shallow feature from $I_{LR}$ called $F_{s} \in \mathbb{R}^{H \times W \times C}$, where $H$ and $W$ represents the height and width of $I_{LR}$ and $C$ is the number of feature's channel.

After that, we employ a stack of multi-scale feature modulation blocks, and then we obtain the deep feature $F_{d} \in \mathbb{R}^{H \times W \times C}$. The multi-scale feature modulation block contains three modules: Efficient Channel Attention Module (ECAM), Multi-Scale feature Fusion Module (MSFM), and Dual-Attention Gated Module (DAGM). MSFM is employed for multi-scale feature extraction, and DAGM is used to integrate spatial and channel-wise features. The MSFN and DAGM are placed behind the ECAM module, and the residual connection is employed to improve the flow of gradients during training. Moreover, we incorporate LayerNorm before the residual connection to facilitate normalization processing and enhance the network's convergence speed.

Finally, we feed the deep features into the image reconstruction block and get the high-resolution SIC image $I_{HR} \in \mathbb{R}^{H \times W \times 1}$. The image reconstruction block is formed by a 3×3 convolution layer for deep features aggregation and pixel shuffle for upscale.

\subsection{Efficient Channel Attention Module (ECAM)}

The details of the ECAM are illustrated in Fig. \ref{fig_1}. It is located before MSFM and DAGM to generate feature map through cross-channel integration, guiding the subsequent deep feature processing modules to focus on more crucial features. Firstly, we use global average pooling to obtain aggregated features, then we utilize two fast 1D convolution layers with the size of 3 to generate channel weights, and then maps the channel weights to [0,1]  by using the Sigmoid activation function. For the input feature $X \in \mathbb{R}^{H \times W \times C}$, the overall formula for calculating channel weights is as follows:

\begin{equation}
    X_{att} = \sigma (\psi_{2}(\psi_{1}(\operatorname{GAP}(X))))
\end{equation}
where GAP is global average pooling, $\psi_{1}$ is the first convolution layer, $\psi_{2}$ is the second convolution layer and  $\sigma$ denotes the Sigmoid activation. After that, we use element-wise multiplication to get the output feature $X_{out}$ as follows:

\begin{equation}
    X_{out} = X \odot X_{att}
\end{equation}

\begin{figure*}[htb]
\centering
\includegraphics [width=6.3in]{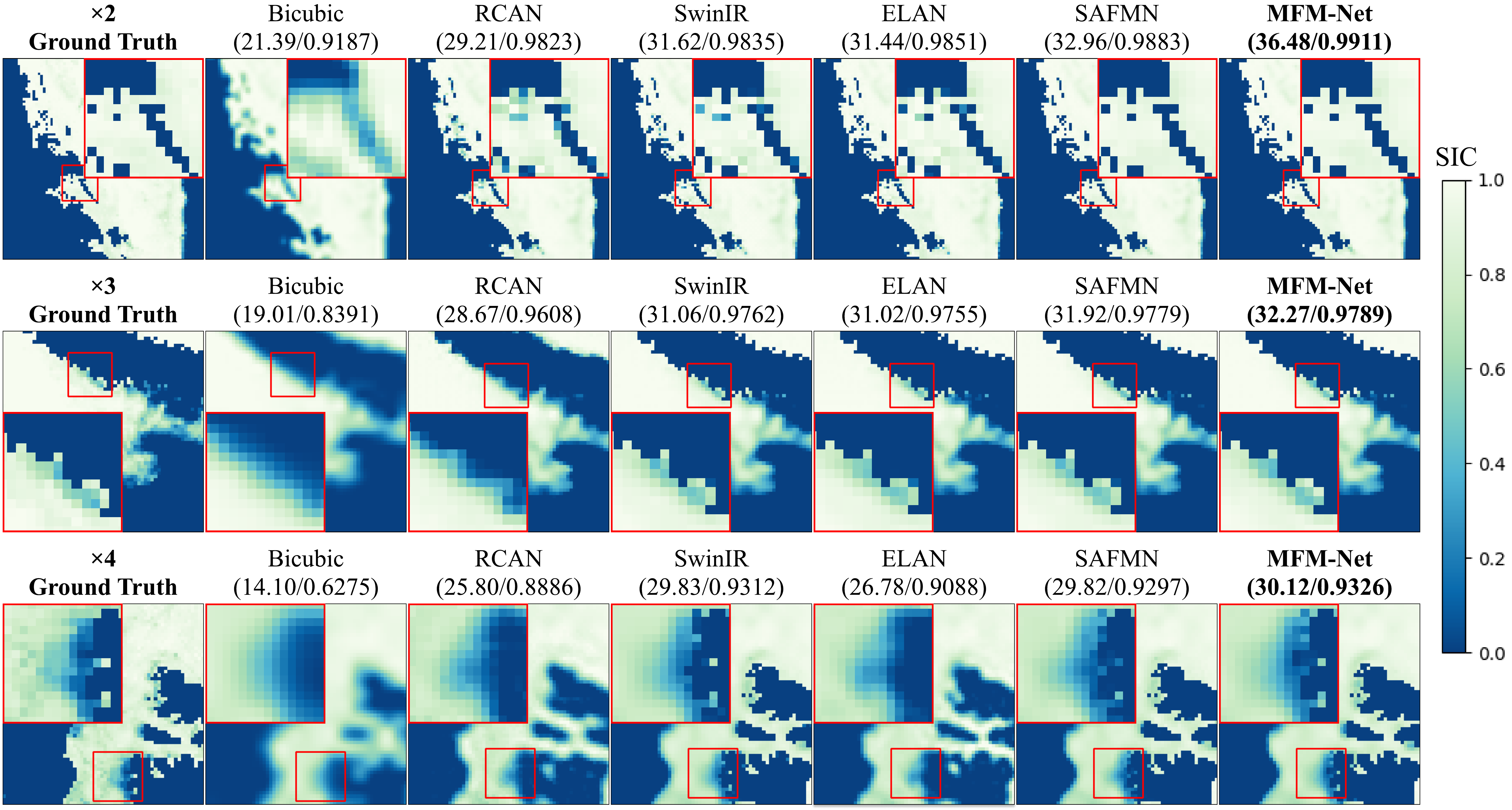}
\caption{Visualization of results of Arctic SIC super-resolution on different methods. The scale of super-resolution for each row is 2, 3, and 4, respectively.}
\label{fig_2} 
\end{figure*}

\subsection{Multi-Scale Feature Fusion Module (MSFM)}

Using large convolution can expand the receptive field, which is beneficial to the super-resolution task. However, only using large convolution kernel will lose some detailed feature information. Therefore, we use convolution kernels of different size to get multi-scale features and retain more information from the input. In addition, we use dilated convolutions with different dilatation coefficient as the large convolution kernel to save computation and improve computational efficiency.

Fig. \ref{fig_1} depicts the details of MSFM. For the input feature $X \in \mathbb{R}^{H \times W \times C}$, we first use channel shuffle operation to enhance feature interaction and get the refined feature $\hat{X} \in \mathbb{R}^{H \times W \times C}$. Then we utilize four branches to perform dilated convolution with the dilate rate of 1, 2, 3, and 4, respectively. Subsequently, we concatenate the four-scale features and use a 1×1 convolution to aggregate the multi-scale features so that we obtain the multi-scale attention map $\overline{X} \in \mathbb{R}^{H \times W \times C}$. After that, we use GELU activation to normalize $\overline{X}$ and then use an element-wise product for $\overline{X}$ and refined feature $\hat{X}$ to get the output feature $X_{out} \in \mathbb{R}^{H \times W \times C}$. The whole procedure can be expressed as follows: 

\begin{equation}
\hat{X} = \operatorname{Channel \_ Shuffle}(X),
\end{equation}
\begin{equation}
\hat{X_i} = \operatorname{Conv}_{3 \times 3, d=i}(\hat{X}), 0\le i \le 3, 
\end{equation}
\begin{equation}
\overline{X} = \operatorname{Conv}_{1 \times 1}(\operatorname{Concat}(\hat{X_0}, \hat{X_1}, \hat{X_2}, \hat{X_3})), 
\end{equation}
\begin{equation}
X_{out} = \hat{X} \odot \phi (\overline{X})
\end{equation}
where $\operatorname{Channel \_ Shuffle}$ is the channel shuffle operation, $\hat{X_i}(0\le i \le 3)$ denotes the output of the four branches, $\operatorname{Conv(\cdot)}$ denotes convolution, $\operatorname{Concat(\cdot)}$ denotes the concatenation operation, $\odot$ denotes the element-wise product and $\phi$ denotes the GELU activation function.

\subsection{Dual-Attention Gated Module (DAGM)}

Due to the homogeneous data structure of Arctic SIC, focusing solely on individual dimensions of features is not conducive to extracting information. Therefore, aggregating features across multiple dimensions is necessary, such as spatial and channel features. DAGM aims to improve the representational ability of the model by extracting and integrating spatial and channel features. The architecture of DAGM is shown in Fig. \ref{fig_1}, which has two main branches to extract the spatial feature $F_s \in \mathbb{R}^{H \times W \times 1}$ and channel feature $F_c \in \mathbb{R}^{1 \times 1 \times C}$ respectively. For the input feature $X \in \mathbb{R}^{H \times W \times C}$, the overall procedure can be expressed as:

\begin{equation}
\hat{X} = \operatorname{Conv}_{3 \times 3}(X),
\end{equation}
\begin{equation}
F_s = \sigma (\operatorname{Conv}_{1 \times 1}(\varphi(\operatorname{Conv}_{1 \times 1}(\hat{X})))),
\end{equation}
\begin{equation}
F_c = \sigma (\operatorname{Conv}_{1 \times 1}(\varphi(\operatorname{Conv}_{1 \times 1}(\operatorname{GAP}(\hat{X}))))),
\end{equation}
\begin{equation}
\overline{X} = (F_s + F_c) \otimes \hat{X},
\end{equation}
\begin{equation}
X_{out} = \operatorname{Conv}_{1 \times 1}(\phi(\overline{X} + X))
\end{equation}
where $\sigma$ denotes the Sigmoid activation function and $\varphi$ denotes the PReLU activation function.

\begin{table}[ht]
\centering
\caption{Super-Resolution Performance of Different Models on the proposed Arctic SIC dataset.}
\vspace{0.5em}
\small
\resizebox{\linewidth}{!}{
\begin{tabular}{rcccccc}
\toprule
\multirow{2}*{Method} & \multicolumn{2}{c}{$\times 2$} & \multicolumn{2}{c}{$\times 3$} & \multicolumn{2}{c}{$\times 4$} \\
\cmidrule(lr){2-3}\cmidrule(lr){4-5}\cmidrule(lr){6-7}
& PSNR & SSIM & PSNR & SSIM & PSNR & SSIM \\
\midrule
\textbf{Bicubic}& 24.74 & 0.8955 & 23.09 & 0.8435 & 22.05 & 0.7977 \\
\textbf{RCAN}  & 32.93 & 0.9695 & 29.55 & 0.9447 & 27.61 & 0.9219 \\
\textbf{SwinIR}  & 33.05 & 0.9703 & 29.73 & 0.9457 & 27.74 & 0.9229 \\
\textbf{ELAN}  & 31.97 & 0.9677 & 28.13 & 0.9372 & 27.30 & 0.9199 \\
\textbf{SAFMN}  & 34.53 & 0.9734 & 29.84 & 0.9461 & 28.01 & 0.9252 \\
\midrule
\textbf{MFM-Net} & 34.58 & 0.9736 & 30.27 & 0.9478 & 28.24 & 0.9370 \\
\bottomrule
\end{tabular}
}
\label{table_1}
\end{table}

\section{Experimental Results and Analysis}

In order to verify the effectiveness of the proposed MFM-Net, extensive experiments are carried out on the Arctic SIC dataset. The dataset used in this paper is generated from the AMSR-E/AMSR-2 SIC data from Key Lab of Polar Oceanography and Global Ocean Change in Ocean University of China. This product is the inversion of daily Arctic SIC based on AMSR-E/AMSR2 orbital brightness temperature data from June 1, 2002 to December 31, 2022, with a spatial resolution of 6.25km. The inversion algorithm used is the Dynamic Tie point ASI algorithm called DT-ASI algorithm, which is developed based on the ASI algorithm of The University of Bremen in Germany.

Due to the vital importance of Arctic SIC at the edge of sea ice for maritime activities such as maritime transportation, we cropped 240 $\times$ 240 pixel data which contain the edge regions from the Arctic SIC data mentioned above as the Ground Truth of the MFM-Net. Then we select 14400 samples from them randomly for training, with 12800 of them used as the training set and the remaining 1600 serving as the validation set

We compare the proposed MFM-Net with some state-of-the-art methods, including bicubic upsample, RCAN \cite{zhang2018image}, SwinIR \cite{liang2021swinir}, ELAN \cite{zhang2022efficient} and SAFMN \cite{sun2023spatially}. The PNSR and SSIM values of all these methods are presented in Table \ref{table_1}. The comparative analysis indicates that the proposed MFM-Net surpasses other methods. The visualization results of super-resolution on different scales are shown in the Fig. \ref{fig_2}, which shows that the proposed MFM-Net can better restore the detailed information of Arctic SIC images than the others methods.

\section{Conclusion}

In this paper, we propose MFM-Net for the Arctic sea ice image super-resolution, which can better extract the deep features of SIC. The ECAM in MFM-Net is used to refine the features of the input. MSFM is used to better aggregate multi-scale features, and DAGM is used for the integration of spatial and channel features. The experimental results on the Arctic SIC dataset from AMSR-E/AMSR-2 SIC data indicates the excellent performance of proposed MFM-Net.

\end{document}